\documentclass[
  twocolumn,
  nofootinbib,
  showpacs,
  ]{revtex4}

\usepackage{graphicx}
\usepackage{amsmath,amssymb,amsfonts}
\usepackage{gensymb}
\usepackage{bm}
\renewcommand{\section}[1]{\par\bigskip\bigskip\textbf{#1} --- }
\newcommand{\eref}[1]{Eq.~\eqref{#1}}
\newcommand{\fref}[1]{Fig.~\ref{#1}}
\renewcommand*{\vec}[1]{\mathbf{#1}}
\newcommand{\MBC}{\ensuremath{(-)}}

\newcommand{\Mpm}{\ensuremath{(\mp,-)}}
\newcommand{\Ab}{\ensuremath{(a,b)}}
\newcommand{\fab}{\ensuremath{f_{\Ab}}}

\newcommand{\fmpm}{\ensuremath{f_{\Mpm}}}

\newcommand{\kab}{\ensuremath{k_{\Ab}}}

\newcommand{\Phiab}{\ensuremath{\Phi_{\Ab}}}

\newcommand{\varthetaab}{\ensuremath{\vartheta_{\Ab}}}

\DeclareMathOperator\erf{erf}
\newcommand{\rdef}{\ensuremath =\mathrel{\mathop:}}
\newcommand{\ddef}{\ensuremath \mathrel{\mathop:}=}
\newcommand{\CCF}{critical Casimir force}
\newcommand{\rmd}{{\rm d}}
\renewcommand*{\div}{\mbox{--}}

\usepackage{bibentry}
\usepackage[T1]{fontenc}
\usepackage{textcomp}
\usepackage{mathptmx}
\usepackage[scaled=0.92]{helvet}
\usepackage{courier}
\begin{document}
\title{\vspace*{2.8cm} Normal and lateral critical Casimir forces between colloids and patterned substrates  \bigskip}
%
%
%
\author{M.~Tr{\"o}ndle}
\author{S.~Kondrat}
\author{A.~Gambassi\footnote{Present address: SISSA - International School for Advanced Studies - and INFN, via Beirut 2-4, 34151 Trieste, Italy}}
\author{L.~Harnau}
\author{S.~Dietrich}
%
%
\affiliation{
  \bigskip
	Max-Planck-Institut f\"ur Metallforschung,  
	Heisenbergstr.\ 3, D-70569 Stuttgart, Germany and\\
	Institut f\"ur Theoretische und Angewandte Physik, 
	Universit\"at Stuttgart, 
	Pfaffenwaldring 57, 
	D-70569 Stuttgart, Germany
  \bigskip
}
\date{November 27, 2009}
%
%
%
\begin{abstract}
	%
	We study the normal and lateral effective critical Casimir forces acting on a spherical 
  colloid immersed in a critical binary solvent and close to a chemically structured 
  substrate with alternating adsorption preference.
  We calculate the universal scaling function for the corresponding potential and compare
  our results with recent experimental data [Soyka~F., Zvyagolskaya~O., Hertlein~C., 
  Helden~L., and  Bechinger~C.,  Phys. Rev. Lett., \textbf{101}, 208301 (2008)].
  The experimental potentials are properly captured by our
  predictions only by accounting for geometrical details of the substrate
  pattern for which, according to our theory, 
  critical Casimir forces turn out to be a sensitive probe.
\end{abstract}
%
\pacs{\\05.70.Jk~~--- Critical point phenomena\\
      82.70.Dd --- Colloids\\
      68.35.Rh --- Phase transitions and critical phenomena\vspace*{1cm}}
\maketitle
%
\section{Introduction.}
The confinement of a fluctuating medium generates effective forces acting on
its boundaries.
A particularly interesting realization of this general principle is provided
by the confinement of concentration fluctuations of a binary liquid mixture
upon approaching a critical demixing point at temperature 
${T=T_c}$ in its bulk phase
diagram~\cite{fisher:1978}.
Generically, the confining surfaces preferentially adsorb one of the two
components of the binary liquid. This amounts to the presence of effective,
symmetry-breaking surface fields favoring either positive $[(+)]$ or negative 
$[(-)]$ values of the scalar order parameter $\phi$ which is the difference 
between the local concentrations of the two species.
The extension of the spatial region in the direction normal to the surfaces, 
within which the local structural properties of the fluid deviate from the bulk 
ones, is given by the bulk correlation length $\xi$, 
which diverges upon approaching 
the critical point as ${\xi(t\to0)=\xi_0^{\pm}|t|^{-\nu}}$.
Here ${t=(T-T_c)/T_c}$ is the reduced temperature\footnote{%
  The two components of a binary liquid mixture are mixed in the disordered 
  phase at $t>0$, whereas a separation in two phases rich in one or the other 
  component occurs in the ordered phase at $t<0$; 
  at \emph{lower} critical points \cite{hertlein:2008,soyka:2008} $t=(T_c-T)/T_c$.
  },
$\nu\simeq0.63$ is a standard bulk critical exponent, and $\xi_0^\pm$ are 
nonuniversal amplitudes for $t\gtrless 0$.
In a film of thickness $L$, the ensuing long-ranged critical fluctuations of
the order parameter lead to a \CCF\ acting on the confining walls \cite{fisher:1978}, 
which is described by a universal scaling function of $L/\xi$ and depends on the type 
of boundary conditions (BC) at the walls (see, e.g., 
Refs.~\cite{krech:9192all,krechbrankov,evans:1994,diehl:2006} and references therein).
This is the thermodynamic analogue of the quantum-electrodynamic Casimir
effect originating from the confinement of vacuum fluctuations
\cite{casimirkardar,gambassi:2009}.
Depending on the relative adsorption preferences of the boundaries, the \CCF\ is either 
attractive for identical symmetry-breaking BC $(\pm,\pm)$ or repulsive for opposite BC $(\pm,\mp)$.
(Symmetry-preserving BC can lead to attractive and repulsive \CCF s, too \cite{krech:9192all,Schmidt:2008}.)
Besides various \emph{indirect} experimental evidences in thin films \cite{allexperiments},
the \CCF\ has been measured also \emph{directly} at the sub-micrometer scale for a colloid 
immersed in a near-critical binary liquid mixture 
close to a homogeneous substrate
\cite{hertlein:2008}. 
Recent Monte Carlo simulations are in quantitative agreement with all available experimental data
\cite{hertlein:2008,huchthasen,vasilyev:0708all}.
\par
Colloids can be used not only as model systems in soft matter physics but also in applications 
on the nano- and micrometer scale which take advantage of their interaction with 
chemically structured solid surfaces.
Such systems can be useful in integrated nano-devices provided that one is able to exert 
active control over these interactions.
Critical Casimir forces provide such a tool, because their \emph{strength} and
\emph{direction} can be tuned via minute temperature changes and surface
treatments of the substrate. 
Recently, the critical Casimir potential of a colloid close to such a chemically patterned substrate
has been measured~\cite{soyka:2008}, providing evidence for the occurrence of 
\emph{lateral} \CCF s.
In Ref.~\cite{soyka:2008}, a dilute suspension of charged spherical colloids, imposing $(-)$ BC to the order parameter of the 
near-critical solvent, is exposed to a chemically patterned substrate, the surface of which consists of alternating stripes, 
which impose $(-)$ and $(+)$ BC. 
The equilibrium spatial distribution of colloids was measured via
digital video microscopy and from it one can define an effective 
potential for a single colloid, which varies laterally due to critical Casimir
forces.
In view of potential applications, surfaces might be designed as to provide temperature-controlled laterally confining
potentials for single colloids, offering novel means of self-assembly 
processes~\cite{soyka:2008}. 
\par
In order to reap the full benefits of this wide range of possibilities, a
thorough theoretical understanding of the underlying physics is essential.
In this context, lateral critical Casimir forces are theoretically known to
occur in the  \emph{film} geometry with  chemically \cite{sprenger} or
geometrically \cite{troendle} structured substrates. However, in the
experimentally relevant realizations of such a film geometry, based on wetting
phenomena \cite{krech:9192all,gambassi:2009} 
at least one of the two confining surfaces (i.e., the
liquid-vapor interface) is laterally 
homogeneous and therefore the lateral critical
Casimir force vanishes. Instead, 
for the geometry of a colloid facing  a substrate 
such a lateral force is expected to arise even if only the substrate
is laterally inhomogeneous, as the experimental findings in
Ref.~\cite{soyka:2008} demonstrate. 
Theoretical studies for the geometry of a colloid facing a wall 
have been limited to \emph{laterally homogeneous} surfaces~\cite{SP:all,hanke:1998}.
In order to overcome these limitations and to interpret, inter alia, the
  experimental data of Ref.~\cite{soyka:2008}, 
we have studied the \CCF\ acting on a sphere
of mesoscopic radius $R$ at surface-to-surface distance $D$ from a substrate
with laterally alternating  adsorption preferences for the two components of a
  confined binary liquid mixture. 
We provide quantitative predictions for the \emph{universal} features of this effective force
(i.e., in excess to regular, nonuniversal background contributions) by using the so-called
Derjaguin approximation (DA) together with the knowledge of the scaling functions determined by
Monte Carlo simulations in spatial dimension $d=3$ for the laterally homogeneous film geometry 
\cite{vasilyev:0708all}.
In order to estimate the accuracy of this Derjaguin approximation, 
we perform a full numerical analysis of the appropriate mean-field theory (MFT) without
further approximation and compare it with the corresponding Derjaguin approximation based on 
analytic results in $d=4$.
%
%
%
\par
The demixing point is approached from the mixed phase by varying the temperature $T$ towards 
$T_c$ at fixed pressure and critical composition of the binary fluid.
Within the field-theoretical renormalization group approach the fixed-point Hamiltonian 
for a binary mixture is given by \cite{binderdiehl}
\begin{equation}
  \mathcal{H}[\phi]=\int_V\,\textrm{d}^d\vec{r}\,\left\{
			 \frac{1}{2}(\nabla\phi)^2
			 +\frac{\tau}{2}\phi^2
			 +\frac{u}{4!}\phi^4
			 \right\},
\end{equation}
where the integration runs over the volume $V$ accessible to the fluid described by the 
$d$-dimensional position vector $\vec{r}$, $\tau \propto t$, and $u > 0$ is a coupling constant. 
In the strong critical adsorption limit \cite{burkh-diehl} the surface contributions to the Hamiltonian 
turn into BC corresponding to infinite surface fields so that $\phi\big|_{\text{surface}}=\pm\infty$. 
Thus, $\mathcal{H}[\phi]$ is supplemented by the BC $\phi=-\infty$ [$(-)$] at the surface of the colloid 
and by $\phi=-\infty$ ($+\infty$) on that part of the substrate with the same (opposite) adsorption preference.
Using a 3$d$ finite element method, we have numerically minimized $\mathcal{H}[\phi=u^{-1/2}\,m]$ and have
obtained the mean-field order parameter profile $m$.
This MFT solution allows one to infer the universal scaling function of the \CCF\ at the 
upper critical dimension $d=4$ up to an overall 
prefactor $\propto u^{-1}$ (and up to logarithmic corrections). 
The normal and lateral \CCF s are calculated using the stress tensor \cite{krech:1997} and the associated 
potential is determined by integration.
%
%
\section{Homogeneous substrate.}
We first consider a three-dimensional sphere with BC $(b)$ facing a homogeneous substrate 
with BC $(a)$, denoting this combination as $(a,b)$. 
The critical Casimir potential $\Phiab$ takes the scaling form\footnote{%
  In ${d=4}$ the $3d$ sphere is a hypercylinder and $\Phiab$ is the potential per 
  length in the fourth direction, along which the physical properties are invariant.}
${\Phiab(D,R,T)=k_BT\,R D^{2-d}\varthetaab(\Theta,\Delta)}$, where $\Delta={D}/{R}$, 
$\Theta=\pm{D}/{\xi_\pm}$ for $t\gtrless 0$, and $\varthetaab$ is a universal scaling 
function~\cite{hanke:1998,hertlein:2008}.
If $D\ll R$, $\varthetaab$ can be expressed -- via the Derjaguin approximation
\cite{derjaguin:1934,hanke:1998,hertlein:2008} -- in terms of the \CCF\ 
per unit area 
$\fab(L,T\gtrless T_c)={k_BT\,L^{-d}\kab(\pm L/\xi_\pm)}$ between two \emph{planar} walls at distance $L$ 
with $(a,b)$ BC.
For $d=3,4$ the Derjaguin approximation gives \cite{hanke:1998,hertlein:2008}
\begin{equation}
\varthetaab(\Theta,\Delta\to0) =2\pi\int_{1}^{\infty}\!\!\textrm{d}s( s^{1-d}-s^{-d})\kab(s\,\Theta). 
\end{equation}
Comparing in $d=4$ the numerically calculated scaling functions $\vartheta_{(\pm,-)}$ to the 
approximate ones obtained via the Derjaguin approximation (on the basis of $k_{(\pm,-)}$ given in Ref.~\cite{krech:1997}), 
we infer that the former are reasonably well described by the latter for $\Delta\lesssim1/3$. 
Assuming a smooth dependence on $d$, we expect this to hold for $d=3$ as well.
%
%
%
\section{Chemically patterned substrate.} 
The essential building block of a substrate with laterally varying adsorption preference is a 
chemical step, i.e., two semi-infinite planes with $(a_\lessgtr)$ BC for $x\lessgtr0$ joined
together, where $x$ is one of the lateral position coordinates;
with respect to the latter the center of the spherical colloid with $(b)$ BC is located at $x=X$.
The resulting critical Casimir potential $\Phi$ for $d=3,4$ 
can be cast in the form
\begin{equation}
  \label{eq:formpotential}
  \Phi(X,D,R,T)         = k_BT\;{R}{D^{2-d}}\;\vartheta\left(\Xi,\Theta,\Delta\right),
\end{equation}
where $\Xi={X}/{\sqrt{RD}}$ is the scaling variable associated with the lateral position of the colloid.
For $\Xi\to\pm\infty$, i.e., sufficiently far from the step, along the
 planar substrate one recovers the homogeneous cases, i.e., 
${\vartheta(\Xi\to\pm\infty,\Theta,\Delta) = \vartheta_{(a_\gtrless ,b)}(\Theta,\Delta)}$.
The \emph{normal} and \emph{lateral} \CCF s acting on the colloid are given by $-\partial_D\Phi$ 
and $-\partial_X\Phi$, respectively.
It is convenient to introduce the scaling function $\omega$ defined via
\begin{equation}
  \label{eq:formtheta}
  \vartheta\left(\Xi,\Theta,\Delta\right)=
  \tfrac{\vartheta_{(a_<,b)} +\vartheta_{(a_>,b)}}{2}
  +
  \tfrac{\vartheta_{(a_<,b)} -\vartheta_{(a_>,b)}}{2} 
  \omega(\Xi,\Theta,\Delta),
  %
\end{equation} 
where $\vartheta_{(a_\gtrless,b)}=\vartheta_{(a_\gtrless,b)}(\Theta,\Delta)$
refer to homogeneous substrates and depend only on $\Theta$
and $\Delta$, while ${\omega(\Xi\to\pm\infty,\Theta,\Delta)=\mp1}$.
Note that the common prefactor $\propto u^{-1}$, which within MFT is left
undetermined by the analytical and numerical determination of
$\vartheta_{(a_\gtrless,b)}(\Theta,\Delta)$ and
$\vartheta(\Xi,\Theta,\Delta)$, does not affect the MFT
prediction for $\omega(\Xi,\Theta,\Delta)$.
\par
For small distances $D\ll R$, i.e., $\Delta\to0$, the surface of the colloid
facing the substrate can be considered to be made up of successive and
consecutive circular rings, parallel to the substrate, which have an
infinitesimal area $\rmd S(l)$ and a radius $r(l)$ which increases with
increasing the normal distance $l$ from the identical corresponding circular
ring obtained by normally projecting the ring on the surface of the sphere
onto the one of the substrate. 
Assuming \emph{additivity} the critical Casimir force $F$ acting on
the colloid is the result of the sum of the forces $\rmd F(l,\ldots)$ acting on
each single pair of such corresponding rings with separation $l$. 
Neglecting also edge effects, $\rmd F(l,\ldots)$ can be expressed in terms
of the critical Casimir force per unit area $f_{(a_\gtrless,b)}(l,T)$ acting
on infinite and \emph{homogeneous} parallel plates with $(a_\gtrless,b)$
BC separated by the same distance $l$ as the rings: 
$\rmd F(l,\ldots) = f_{(a_>,b)}(l,T)\rmd S_> + f_{(a_<,b)}(l,T)\rmd S_<$.
Here $\rmd S_\gtrless$ (with $\rmd S(l) = \rmd S_> + \rmd S_<$) indicates the
surface area 
of that portion of the ring on the substrate for which 
$x\gtrless 0$, i.e.,
corresponding to $(a_\gtrless)$ BC. This area $\rmd S_\gtrless$ depends, inter
alia, on $l$ and on the position of the colloid. For the potential $\Phi$ (see
\eref{eq:formpotential}) associated with $F$ we eventually find in
$d=3,4$ %
\begin{equation}
\begin{split}
  \label{eq:steppot}
  &\vartheta(\Xi\gtrless0,\Theta,\Delta\to0)=
  \vartheta_{(a_\gtrless,b)}(\Theta,\Delta\to0)
  \pm
  \tfrac{1}{2}{\Xi^4}\\\times
  &\int\limits_{1}^{\infty}\textrm{d}s 
   \frac{s\, \arccos\left(s^{-1/2}\right)-\sqrt{s-1}}{\left( 1+{\Xi^2s}/{2} \right)^d}
   \Delta k \left( \Theta\left(1+{\Xi^2s}/{2}\right)\right),
\end{split}
\end{equation}
where $\Delta k (\Theta) = k_{(a_<,b)}(\Theta) - k_{(a_>,b)}(\Theta)$ is the difference between 
the scaling functions of the \CCF s acting on two planar, homogeneous walls with $(a_<,b)$ 
and $(a_>,b)$ BC, respectively.
Equation~(\ref{eq:steppot}) yields $\omega(\Xi=0,\Theta,\Delta\to0) = 0$ (expected from \eref{eq:formtheta} 
and additivity) and 
\begin{equation}
  \label{eq:da-crit}
  \omega(\Xi,\Theta=0,\Delta\to0) = 
  {\Xi\left(1-d-\Xi^2\right)}{\left(\Xi^2+2\right)^{-3/2}},
\end{equation}%
\emph{independent} of $k_{(a_\gtrless,b)}$.
Note that Eqs.~(\ref{eq:formpotential})--(\ref{eq:da-crit}) are valid beyond the BC we consider in the 
following, i.e., $a_\gtrless\in\{+,-\}$ and $b = -$. 
For $(\mp,-)$ BC the critical Casimir force $\fmpm(D,T)$ between two planar
walls decays $\propto\exp(-\Theta)$ as a function of $\Theta$ 
for $\Theta\gg1$
\cite{evans:1994,borjan:2008,krech:1997}\footnote{%
  This purely exponential decay is proven analytically 
  only in $d=2$ and $4$ but 
  it is expected to hold for all spatial dimensions.}%
, which leads to a \emph{$d$-independent} 
result for $\omega$:
\begin{equation}
  \label{eq:erf}
  \omega(\Xi,\Theta\gg1,\Delta\to0) = 
 - \erf (\sqrt{\Theta/2}\;\Xi),
\end{equation}
where $\erf$ is the error function.
%
\begin{figure}
	\includegraphics[trim=0 2 0 0,clip]{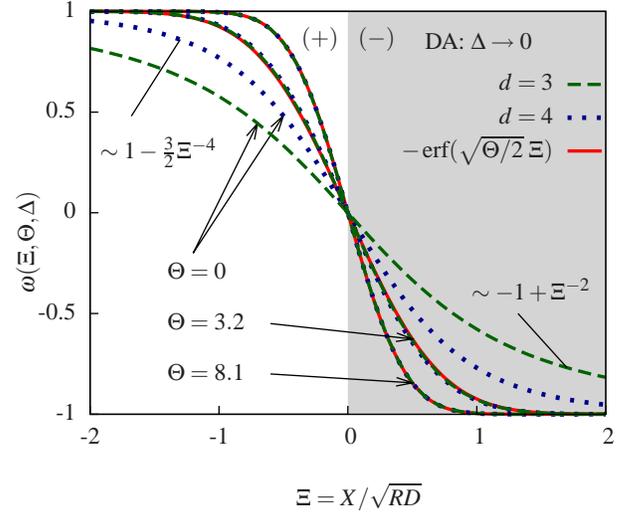}
  \caption{ 
    (Color online)
    Scaling function $\omega$ [\eref{eq:formtheta}] within Derjaguin approximation (i.e.,
    $\Delta\to0$) for the critical Casimir potential of a sphere $(-)$
    facing a chemical step $(+|-)$.
    The dashed and dotted lines refer to $d=3$ and $d=4$ (MFT),
    respectively.
    At bulk criticality ($\Theta=0$), $\omega$ is given by \eref{eq:da-crit}, 
    whereas for $\Theta\neq0$ it is calculated on the basis of the
    scaling functions for the film geometry [see the main
    text].
    For $\Theta\gtrsim3$, $\omega$ becomes practically
    independent of $d$ and coincides with the expression for $\Theta\gg1$
    [\eref{eq:erf}, solid lines, barely distinguishable from the corresponding
    dashed and dotted ones].
  }
  \label{fig:compare}
\end{figure}
\par
\begin{figure}
	\includegraphics[trim=0 2 0 0,clip]{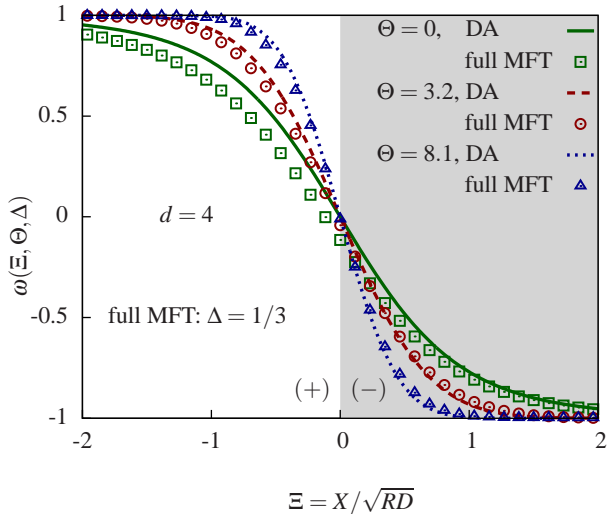}
  \caption{
    (Color online)
    Scaling function $\omega$ [\eref{eq:formtheta}] in $d=4$ (MFT)
    for the critical Casimir potential of a sphere $(-)$ facing a chemical step $(+|-)$.
    The numerical results for $\Delta=1/3$ are indicated by symbols, the size
    of which refers to the  numerical error.
    The lines show the corresponding results for $\Delta\to0$, calculated
    within the Derjaguin approximation, 
    which provides a good approximation to the numerical data for
    $\Theta\gtrsim3$.
    }
  \label{fig:damft}
\end{figure}

In \fref{fig:compare} we compare the behavior of $\omega$ calculated within the Derjaguin approximation in $d=4$ and $3$; 
the required scaling functions $k_{(\pm,-)}$ for the film geometry [see \eref{eq:steppot}] are 
obtained analytically in $d=4$ within MFT \cite{krech:1997} and  in $d=3$ from
Monte Carlo simulations \cite{vasilyev:0708all}.
The systematic uncertainty of the latter does not affect significantly (at most by $3\%$) the estimate 
of $\omega$ shown in \fref{fig:compare}.
For $\Theta\to0$ the critical Casimir potential for $d=3$ as a function of the lateral coordinate varies more 
smoothly than the corresponding MFT function [see \eref{eq:da-crit}].
However, for $\Theta\gtrsim3$ the scaling functions $\omega$ for $d=3$ and $d=4$ (MFT) practically 
coincide with \eref{eq:erf} valid for $\Theta\gg1$.
Figure~\ref{fig:damft} compares the scaling function $\omega$ for $d=4$ (MFT) calculated in the limit 
$\Delta\to0$ (Derjaguin approximation) with the full one determined numerically for $\Delta=1/3$.
The former provides a very good approximation of the latter, especially for
large $\Theta$. 
Since a viable strategy for carrying out Monte Carlo simulations for the pertinent lattice Ising model in the sphere-plate 
geometry is currently not available, this study of the DA in $d=4$ is an important piece of information because at present this
is the only way to quantitatively assess the range of parameters within which the DA provides an accurate approximation 
of the actual scaling function. 
On the basis of these results in $d=4$ and assuming a smooth dependence on $d$, we expect that also in $d=3$ the 
DA captures the actual behavior of the scaling function for small $\Delta$ and large $\Theta$ ---two conditions 
which are met by the experimental data discussed below.
%
\par
A consequence of the underlying assumption of additivity is that 
the critical Casimir potential  due to an arbitrarily striped substrate can be calculated within the 
Derjaguin approximation on the basis of \eref{eq:steppot} as the appropriate superposition of consecutive, 
shifted steps.
From the comparison of the Derjaguin approximation for this potential in $d=4$ with the corresponding MFT 
results we find that the former describes quite well the actual behavior predicted by the latter
even for substrates with a fine pattern, at least as long as the ratios between the
characteristic lengths of the pattern and the geometrical average $\sqrt{RD}$ of the radius $R$ of the colloid and 
its distance $D$ from the substrate are not too small.
In particular for a chemical \emph{stripe} of finite width $L\gtrsim0.2\sqrt{RD}$ the use of the Derjaguin approximation for $\Delta\lesssim1/3$
is justified. 
Moreover, for $L/\sqrt{RD}\gtrsim3$ and $\Theta\gtrsim3$ or for $L/\sqrt{RD}\gtrsim2$ and $\Theta\gtrsim8$ the stripe is 
almost equally well described by two independent and subsequent chemical steps.
Similarly for \emph{periodically} patterned substrates of alternating stripes with total period $P$ the DA describes 
the actual data accurately if, in addition to the fulfillment of the previous conditions for each single stripe in the pattern,
one has $P\gtrsim3\sqrt{RD}$.
We expect that this property carries over to the case of $d=3$.

%
%
\break
\section{Comparison with the experiment by Soyka \textit{et al.}} 
In Ref.~\cite{soyka:2008} the substrate surface confining the colloidal suspension 
consists of stripes along the $y$-axis which impose $(-)$ and $(+)$ BC alternating along
the $x$-axis, and have a width $L_-= 2.6\,\mu$m and $L_+ = 5.2\,\mu$m, 
resulting in a periodicity $P=L_-+L_+= 7.8.\,\mu$m.
\par
The total potential $\Phi_{\rm tot}$ (in units of $k_BT$ with $T\simeq T_c \simeq 307\,$K)
of the forces acting on each colloid ($R=1.2\,\mu$m, \MBC  BC) is the sum of the electrostatic, gravitational, and critical Casimir contributions:
\begin{equation}
  \label{eq:phitot}
  \Phi_{\rm tot}(x,D) = \Phi_{\rm el}(x,D) + \Phi_{\rm g}(D) + \Phi_{\rm C}(x,D).
\end{equation}
In general the colloidal particle interacts with the substrate and the surrounding medium also via van der Waals 
forces\cite{dantchev:2007,hertlein:2008}, the potential of which should be added to the rhs of Eq.~\eqref{eq:phitot}.%
\footnote{%
Upon approaching the critical point, the temperature-dependent dielectric permittivity $\epsilon(T)$ of the binary liquid mixture exhibits a 
weak cusplike singularity 
$\epsilon(T)-\epsilon(T_c) \propto |t|^{1-\alpha}$, 
where $\alpha\simeq 0.11$ is the critical exponent of the specific heat for the three-dimensional Ising universality 
class~\cite{sengers}. 
This weak variation of $\epsilon(T)$ might affect the strength of the van der Waals forces, as well as the range of 
the electrostatic interaction $\Phi_{\rm el}$~\cite{hertlein:2008}.
However, in the near-critical mixture of water and lutidine, the permittivity $\epsilon(T)$ turns out to vary less 
than $1\%$ for $|T_c-T|<1\,$K~\cite{kaatze}, and therefore the van der Waals forces as well as the electrostatic 
interaction are expected to be not affected significantly.  %
}%
However, it has been shown that for the particular choice of materials and experimental conditions used in 
Ref.~\cite{soyka:2008}, the corresponding contribution is negligible compared with the others \cite{hertlein:2008,sprenger}.
For the equilibrium number density of colloids one has $\rho(x,y,D)\propto \exp\{-\Phi_{\rm tot}(x,D)\}$, 
where $x$ and $y$ are the coordinates of the projection of the center of the colloid onto the substrate surface. 
Since the distance $D$ is not resolved in the setup of Ref.~\cite{soyka:2008}, only the 
projected number density $\int_0^\infty\!\!{\rm d}D\;\rho(x,y,D)$ is experimentally accessible. 
Due to the translational invariance of the pattern along the $y$-direction (of length $\ell \gg R,L_\pm$) 
one can in addition project the experimental data onto the $x$-axis, 
$\ell^{-1}\!\int_0^\ell\!{\rm d}y\int_0^\infty\!\!{\rm d}D\;\rho(x,y,D) \rdef \hat\rho(x)$, 
and define, up to a constant, an effective potential $\hat V$ such that
$\hat\rho(x)\propto\exp\{-\hat V(x)\}$~\cite{soyka:2008}. 
Our analysis shows that for the experimental conditions used in 
Ref.~\cite{soyka:2008} the effects of consecutive chemical steps do not interfere and therefore 
one can focus on a \emph{single step} located at $x=0$, with $(\mp)$ BC for $x \gtrless 0$. 
At this stage we assume that each chemical step experimentally realized on the
substrate is effectively sharp and straight at the micrometer scale of the
problem. %
For such a step, $\Phi_{\rm el}(x,D)$ interpolates between the screened electrostatic potentials 
$\Phi_{\rm el,\mp}(D)$ of a colloid facing a homogeneous substrate, which are recovered far from 
the step for $x\gtrless 0$ and which are well approximated by $\exp\{-(D-D^\mp_0)/\lambda\}$ 
where $\lambda \simeq 12\,$nm is the screening length of the mixture and 
$D_0^\mp \simeq 0.1\div0.2\,\mu$m~\cite{hertlein:2008}.
Within the Derjaguin approximation
\begin{equation}
  \Phi_{\rm el}(x,D) = \Phi_{\rm el,+}(D)\theta(-x/\Lambda) + \Phi_{\rm el,-}(D)\theta(x/\Lambda), 
\end{equation}
where $\theta(u) = [1+\erf(u)]/2$ and $\Lambda = \sqrt{2 R \lambda} \simeq 0.17\,\mu$m. 
The gravitational potential in units of $k_BT$ at $T\simeq307$K 
is ${\Phi_\textrm{g}(D)=GD}$ with 
$G = (\rho_{\rm PS}-\rho_{\rm WL})(4\pi R^3/3)g /(k_BT) \simeq 1.12\,
(\mu\mbox{m})^{-1}$ for a gravitational acceleration $g=9.8\,\mbox{m/s}^2$, where $\rho_{\rm PS} \simeq 1.055\,\mbox{g/cm}^3$ and
$\rho_{\rm WL} \simeq 0.988\,\mbox{g/cm}^3$~\cite{Jayalakshmi:1994} are the mass densities of the polystyrene 
colloid and the solvent, respectively.
$\Phi_C$ is calculated on the basis of Eqs.~(\ref{eq:formpotential}) and~(\ref{eq:steppot}) with 
$\vartheta_{(\pm,-)}$ given by the Derjaguin approximation for homogeneous substrates and $k_{(\pm,-)}$ obtained 
from Monte Carlo simulations~\cite{vasilyev:0708all}. 
Far from the critical point $\Phi_{\rm C}$ is negligible compared to $\Phi_{\rm el,g}$ and the average height 
$\langle D \rangle$ of the colloid above the substrate is $\simeq D_0^\mp + G^{-1} \simeq 1\, \mu\mbox{m}$, 
with typical fluctuations $\simeq G^{-1} \simeq 0.8\,\mu$m. 
(These values are significantly larger than those
  reported in Ref.~\cite{soyka:2008}, which were extrapolated from the
  different experimental conditions used in Ref.~\cite{hertlein:2008}.)
Upon approaching the critical point, the repulsive \CCF\ for $x<0$ pushes the colloid slightly away from the substrate, 
causing only a minor increase of $\langle D \rangle$. 
For $x>0$, instead, as soon as the correlation length $\xi$ exceeds a certain $D_0^-$-dependent value, the colloid abruptly localizes in 
the potential well due to the interplay between the attractive critical
  Casimir force and the electrostatic repulsion, 
with $\langle D \rangle \simeq D_0^-$ and fluctuations of few tens of nm.
As a result, at distances $D \simeq \langle D\rangle$, the contributions of $\Phi_{\rm el}$ to $\Phi_{\rm tot}(x<0,D)$ 
and of $\Phi_{\rm g}$ to  $\Phi_{\rm tot}(x>0,D)$ are negligible and the behavior of the colloid depends 
sensitively only on the actual value of $D_0^-$.
Accordingly, we fix $D_0^+=0.1\,\mu$m, $\lambda=12\,$nm and calculate $\hat V(x)$ and $\Delta\hat V \ddef 
\hat V(-L_+/2) - \hat V(L_-/2)$ on the basis of $\Phi_{\rm tot}(x,D)$ for the single step. 
$\hat V$ and $\Delta \hat V$ depend on $D_0^-$ and, via the bulk correlation length 
$\xi\simeq\xi^+_0[(\Delta T_c + \Delta T)/307\,\mbox{K}]^{-0.63}$,  
on the estimated distance $\Delta T=T_c-T$ from the critical point (actually located at $T = 307\mbox{K} + \Delta T_c$). 
The values $D_0^-\simeq 0.136\,\mu$m, $\xi^+_0\simeq 0.42\,$nm, and $\Delta T_c \simeq 27\,$mK yield a very good fit 
to the experimental data for $\Delta\hat V$ (Fig.~3 of Ref.~\cite{soyka:2008}).
Whereas $\Delta T_c$ is within the experimental accuracy, $\xi^+_0$ is significantly larger than previous estimates
$\bar\xi^+_0=(0.2\pm0.02)\,$nm~\cite{hertlein:2008}, suggesting that for the rather small corresponding values 
of $\xi\simeq 20\div 36\,$nm corrections to the leading scaling behavior might still be relevant. 
For $x<0$ the contribution of $\Phi_{\rm C}$ to $\Phi_{\rm tot}$ is
significant only for distances from the substrate 
$D\lesssim \xi$, which corresponds to 
$\Delta = D/R \lesssim 0.03$, whereas for $x>0$ the typical distance is $D\simeq D_0^-$ and therefore $\Delta \simeq 0.12$ with
$\Theta = D/\xi \gtrsim 4$.
In addition, the relevant geometrical parameters are $P/\sqrt{RD}\simeq7\div20$ and $L_-/\sqrt{RD}\simeq2\div7$,
such that, based on our theoretical analysis of the range of validity of the DA, we expect the latter to be accurate 
for the potentials $\Phi_{\rm el}$ and $\Phi_{\rm C}$.
Moreover, the values of $\Theta$, $P/\sqrt{RD}$, and $L_-/\sqrt{RD}$ are such that the
resulting critical Casimir potential is adequately described by the superposition of a sequence of single chemical steps.
%
\begin{figure}
    \includegraphics[trim=0 2 0 0,clip]{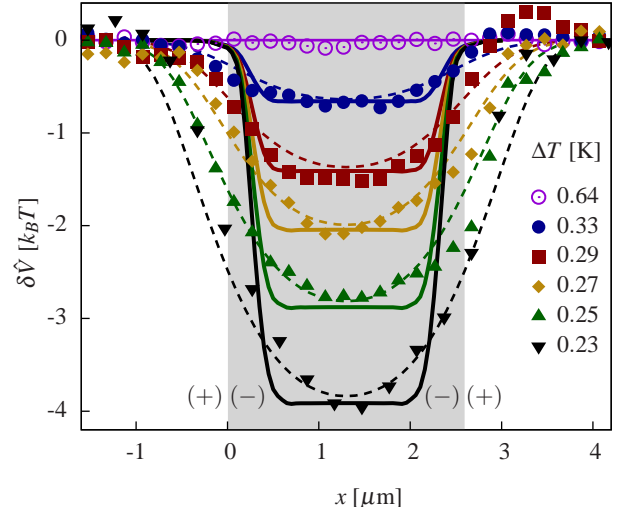}
  \caption{
    (Color online)
    Lateral variation of the effective potential $\delta\hat V$ [see the main text] of
    a colloidal particle $(-)$ facing a  chemically patterned substrate and  immersed in a binary 
    liquid mixture at critical concentration for various temperatures $T_c-\Delta T$. 
    Symbols indicate the experimental data of Ref.~\cite{soyka:2008}, whereas the solid and the dashed 
    lines are theoretical predictions for the same values of parameters for an ideal and non-ideal 
    stripe pattern, respectively, based on \protect{\eref{eq:steppot}}.
	}
  \label{fig:comp-exp}
\end{figure}

\par
In Fig.~\ref{fig:comp-exp} we compare the experimental data with the resulting theoretical predictions for 
$\delta\hat V(x)\ddef\hat V(x)-\hat V(-L_+/2)$ (solid lines) across the $(-)$ stripe (grey). 
As anticipated, the effects of the chemical steps at $x=0$ and $L_-$ do not interfere in the actual range
of parameters. 
The theoretical curves -- in \emph{qualitative} disagreement with the experimental data -- display a sharp 
transition between the plateau values $0$ and $- \Delta \hat V$.
In order to test the robustness of this distinctive 
feature we varied  $\lambda$ within the plausible range $8\div 18\,$nm, 
considered polydispersity ($R=1.2\div 1.8\,\mu$m), and allowed for a possible inhomogeneous buoyancy 
$\Phi_\textrm{g}(D,x)$ induced by laterally varying fluid layers adsorbed on the colloid and on the substrate. 
Moreover, due to their fabrication process, the $(-)$ stripes might have a rather weak preferential adsorption as 
compared to the $(+)$ stripes, which we tried to capture by reducing the amplitude of $k_{(-,-)}$ by up 
to 70\%.
Consequently, the values of $D_0^-\simeq 0.10\div0.14\,\mu$m, $\xi^+_0\simeq 0.3\div 0.4\,$nm, 
and $\Delta T_c\simeq -0.1\div 0.1\,$K, which yield the best agreement with the experimental data for $\Delta \hat V$, 
are affected by these changes\footnote{
  As the amplitude of $k_{(-,-)}$ is reduced, the fitted value of $\xi^+_0$ moves
  closer to $\bar\xi_0^+$,  suggesting that indeed the preferential adsorption
  of the $(-)$ stripes might be effectively rather weak.
  }
but the sharpness of the variation \emph{is not}.
In addition, our analysis shows that the effect of the periodic spatial arrangement of the stripes -- which would 
smoothen the potential for relatively small periodicities $P$ -- is negligible for the experimental conditions 
used in Ref.~\cite{soyka:2008}.
On the same basis, we expect the DA to be sufficiently accurate and that the
non-linearities inherent in the critical Casimir interaction, which actually invalidate the assumption of additivity 
of the forces and might cause a smoothening, \emph{do not} affect significantly the potential shown in 
\fref{fig:comp-exp}.
\par
However, if the actual position $x=x_s(y)$ of each chemical step varies smoothly along the $y$-axis on the length scale 
$\ell \simeq 40\,\mu$m explored by the colloid during the measurement, the projection of the number density $\rho$ 
onto the $x$-axis results in a smoothed distribution $\hat \rho(x)$. 
Even though there is no direct measurement of such a variation of the position of the step,
it is reasonable to assume that it occurs on the length scale $\ell$ due to the fabrication 
process (focused ion beam acting on glass) and due to the projection of the data onto one 
dimension without independent knowledge of the alignment of the chemical stripes.
In order to estimate the consequences, 
we have assumed the total potential of the forces to be given by
$\Phi_{\rm tot}(x-x_s(y),D)$ with $x_s(y)$ 
characterized by a Gaussian distribution $p(x_s)$ with zero average and standard deviation $\Delta x = 0.5\,\mu$m. 
($\Delta x$ may contain a contribution from a smooth intrinsic chemical gradient.)
Thus, for the number density, the projection $\ell^{-1}\! \int_0^\ell\!{\rm d}y$ turns into 
$\int_{-\infty}^{+\infty}\!{\rm d}x_s\, p(x_s)$.
The resulting $\delta\hat V$, 
which basically correspond to a convolution of the original
almost square-well-like potentials with a Gaussian of width $\Delta x$, 
are shown as dashed curves in Fig.~\ref{fig:comp-exp}.
The agreement with the experimental data is significantly improved.
In view of the pronounced modifications of the resulting potential our analysis demonstrates 
that \CCF s respond sensitively to geometrical details of the chemical pattern, which could not 
be checked independently in Ref.~\cite{soyka:2008}. 
\par
We conclude that our theoretical analysis provides a quantitative
understanding of lateral critical Casimir forces and new insights into the
corresponding experiments. The reliability of the theoretical description
allows one to use critical Casimir forces for probing microscopic features of
the system which are difficult to access otherwise.
%
%
\begin{center}
  ***
\end{center}
We thank C.~Bechinger, L.~Helden, C.~Hertlein, U.~Nellen,
 F.~Soyka, and O.~Zvyagolskaya for very useful discussions.
SK and LH gratefully acknowledge support by the grant HA~2935/4-1 of the Deutsche Forschungsgemeinschaft.
%
%
\newpage
\end{document}